# Formation of Copper oxide II in polymer solution-blow-spun fibers and the successful non-woven ceramic production

Alex Nascimento Bitencourt da Silva, Marcia Regina de Moura, Rafael Zadorosny*

*Universidade Estadual Paulista (UNESP), Faculdade de Engenharia de Ilha Solteira, Departamento de Física e Química*

*rafael.zadorosny@unesp.br

**Abstract**
*Copper oxide II* is a p-type semiconductor that can be used in several applications. Focusing on producing such material using an easy and low-cost technique, we followed an acetate one-pot-like route for producing a polymer precursor solution with different acetates:PVP (polyvinylpyrrolidone) weight ratios. Then, composite nanofibers were produced using the solution blow spinning (SBS) technique. The ceramic CuO samples were obtained after a calcination process at 600ºC for two hours, applying a heating rate of 0.5ºC/min. Non-woven fabric-like ceramic samples with average diameters lower than 300 nm were successfully obtained. SEM images show relatively smooth fibers with a granular morphology. XRD shows the formation of randomly oriented grains of CuO. In addition, FTIR and XRD analyses show the CuO formation before the heat treatment. Thus, a chemical reaction sequence was proposed to explain the results.

## 1. INTRODUCTION

The copper oxide II (CuO) is a transition metal oxide with the characteristic of being a p-type semiconductor showing a narrow bandgap in the infrared region (1.2 eV) [1]. Besides, CuO presents a high capacity to absorb solar radiation and high thermal conductivity (33 W/mK) [2]. These characteristics make it promising for use as photodetector cells [3], gas sensors [4], and inorganic pollutant removers [5]. Also, CuO is the base for some types of high critical temperature superconductors (HTSC). CuO can be found at the nanoscale in different sizes and morphologies, such as in 0D nanoparticles, 1D nanowires, 2D rods, and 3D structures such as nanoflowers [6-9]. These different morphologies can be controlled and tailored during the material's synthesis process.

Several traditional methods have been used to produce CuO nanostructures, such as the hydrothermal/solvothermal method. This is a common method to effectively produce CuO micro or nanostructures due to its easy control of the final materials' sizes and morphologies [10]. Other methods can also be used for the synthesis of such a material, such as electrochemistry [11], chemical precipitation [12], and thermal conversion of solid precursors [13], among others.

New studies focusing on CuO nanofibers have paid attention to their photocatalytic ability to degrade organic pollutants [14]. In such a case, CuO is generally used as a co-catalyst since the combination of its narrow bandgap with the large bandgap presented by catalyst materials (such as ZnO [15] and $TiO_2$ [16]), improve the photocatalytic activity of solar cells. As non-woven nanofiber samples are naturally porous materials with a high surface-to-volume ratio, research interest for applying CuO composites as gas sensors [17] has been raised. Combining the described nanofibers' properties can allow future sensors' production with faster recovery, and better sensitivity and response time [17].

In this work, we report CuO nanowires' production using a one-pot-like methodology to synthesize precursor solutions combined with the Solution-Blow-Spinning (SBS) technique. The one-pot-like methodology, reported by Rotta et al. [18], was used as an alternative chemical route to produce polymer and ceramic materials with high stability, low cost, and reduced number of steps. The method is based on carrying out the various chemical reactions necessary to obtain the final material in a single vessel. Besides, the SBS technique was chosen due to its high production rate [19], ability to form long nanofibers (of the order of micrometers) with excellent interconnectivity between them, and high porosity. Such characteristics are useful for applications such as, e.g., photocatalytic materials [14]. In the preparation of the precursor solution, polyvinylpyrrolidone (PVP) was used as a polymer matrix. PVP also presents reducing and stabilizing properties, characteristics of its hydroxyl groups (OH-), making it an ideal reducer for the production of metallic nanoparticles [20]. In addition, PVP has been used to produce non-woven YBCO ceramic nanofibers by SBS [21,22].

## 2. EXPERIMENTAL PROCEDURES
### 2.1. MATERIALS

The precursor polymer solution was prepared using the reagents as follows. Copper acetate monohydrate [$Cu(CH_3CO_2)_2 \cdot H_2O$] (99%), polyvinylpyrrolidone (PVP, MM = 1300,000 g/mol), glacial acetic acid, propionic acid, and methanol. All products were purchased in analytical grade from Sigma Aldrich, and the quantities used will be described in the next section.

### 2.2. PREPARATION OF THE PRECURSOR POLYMER SOLUTION

The method used here to prepare the precursor solution was based on the work published by Rotta et al. [21]. Nonetheless, to obtain a homogeneous solution without precipitates, the proportions between the used solvents had to be changed from [21] to 5v/v% acetic acid, 50v/v% propionic acid, and 45v/v% methanol, totaling 16 mL. The increase in the proportion of propionic acid used here is justified in that only copper acetate II was used, and such a solvent is responsible for CuO solubilization. In previous work, Rotta et al. [21] described that the weight ratio between acetates and PVP (Ac:PVP) is crucial for obtaining a precursor solution with a viscosity adequate to be used in the SBS apparatus. In the present work, the variation of the Ac:PVP ratio ranged from 1:1 to 7:1. However, we present some results obtained from samples prepared by 5:1 and 7:1 Ac:PVP ratios, which are labeled as S51 (for samples produced using Ac:PVP = 5:1) and S71 (following the same idea).

Figure 1 schematically shows the one-pot method used to prepare the precursor solution, and Table 1 shows the reagents' quantities used to prepare S51 and S71 samples. The copper acetate was weighed to obtain 2 g of CuO, and then it was placed in a vessel. Thus, under magnetic stirring, 5 v% of acetic acid, 50 v% of propionic acid, and 45 v% of methanol were added precisely in the mentioned sequence. It is worth mentioning that this volume percentage may change depending on the solubility of the salt used. Just after the solvents' addition, the beaker was tightly closed to avoid solvent evaporation. After 5 min stirring, the PVP was added slowly, and the vessel closed in sequence. The solution was then left under constant stirring for 24 hours to assure its stability. The entire process was carried out at room temperature (of about 26-28ºC), resulting in a dark blue solution without precipitates.

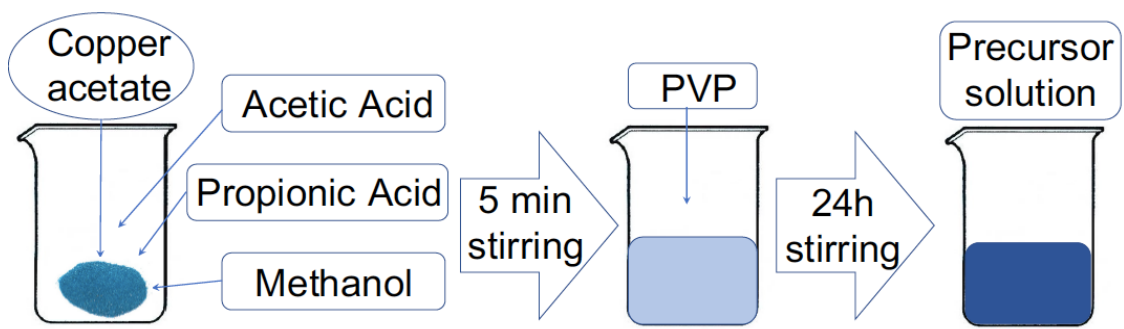

**Figure 1:** Schematic diagram of the one-pot-like route used to prepare the precursor solution.

**Table 1** – Reagents and the quantities used to prepare the precursor solution with Ac:PVP ratios of 5:1 and 7:1.

|     | Copper acetate (g) | Acetic acid (mL) | Propionic acid (mL) | Methanol (mL) | PVP (g) |
| --- | --- | --- | --- | --- | --- |
| **5:1** | 5.07 | 1.0 | 10.1 | 9.1 | 1.01 |
| **7:1** | 6.34 | 0.9 | 9.1 | 8.1 | 0.91 |

### 2.3. VISCOSITY

The precursor solution's viscosity is of fundamental importance for uniform fibers' formation in the SBS technique. Such a parameter is directly related to the polymer concentration used in the solution [23]. When the solution's viscosity is very low, the blowing process creates unstable jets, resulting in inhomogeneous fibers with many drops in their extension [24]. By increasing the polymer concentration in solution, the viscosity increases, and the blowing process can occur without instabilities and beads formation. However, a critical PVP concentration must be analyzed since it influences the final nanowire's morphology [23]. Based on previous works, a concentration of 5wt/v% of the polymer poly(vinylpyrrolidone) (PVP) was used in the precursor solution in the present work [21,25].

The seven precursor solutions studied here were analyzed from the point of view of their rheological behavior. Figure 2 shows the viscosity as a function of the Ac:PVP ratio. It is worth mentioning that the ratios 1:1, 2:1, 3:1, and 4:1 produced jet instabilities making the blowing process inadequate to produce continuous and homogeneous nanowires. By increasing the Ac:PVP ratio, e.g., 5:1, 6:1, and 7:1, the solutions' viscosity increased, and the blowing process was facilitated, generating continuous and smooth nanowires without beads.

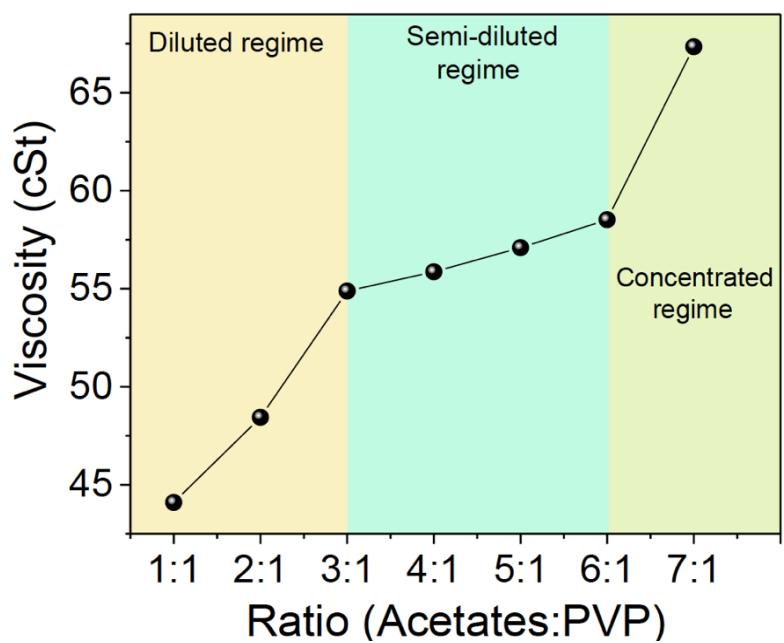

**Figure 2:** Viscosity as a function of the Ac:PVP ratio. The diluted, semi-diluted, and concentrated regimes can be identified, following the classification in Ref. [21].

The increased viscosity with the Ac:PVP ratios can be related to the entanglement of the PVP networks with the $Cu^{2+}$ cations. One can also note in Figure 2 the three regions reported in Ref. [23], i.e., from 1:1 to 3:1 is a diluted regime, from 3:1 to 6:1 there is the semi-diluted regime, and above 6:1 the concentrated regime takes place. Besides that, the average diameter of the nanowires decreases for lower ratios and there is difficulty in producing continuous and homogeneous wires.

## 2.4. OBTAINING THE POLYMER SAMPLES VIA SBS

Non-woven CuO was obtained by the SBS technique following the same process as that described in Ref. [21]. However, some SBS parameters were modified to adjust to the used solution, i.e., the solution injection rate was set to 50 µL/min, the inner needle's diameter was 0.50 mm (25G), the working distance was 35 cm, the angular velocity of the collector was 40 rpm, and the compressed air pressure was adjusted to 0.6 bar. Using the SBS technique with the described parameters, a non-woven polymer was obtained, as shown in Figure 3.

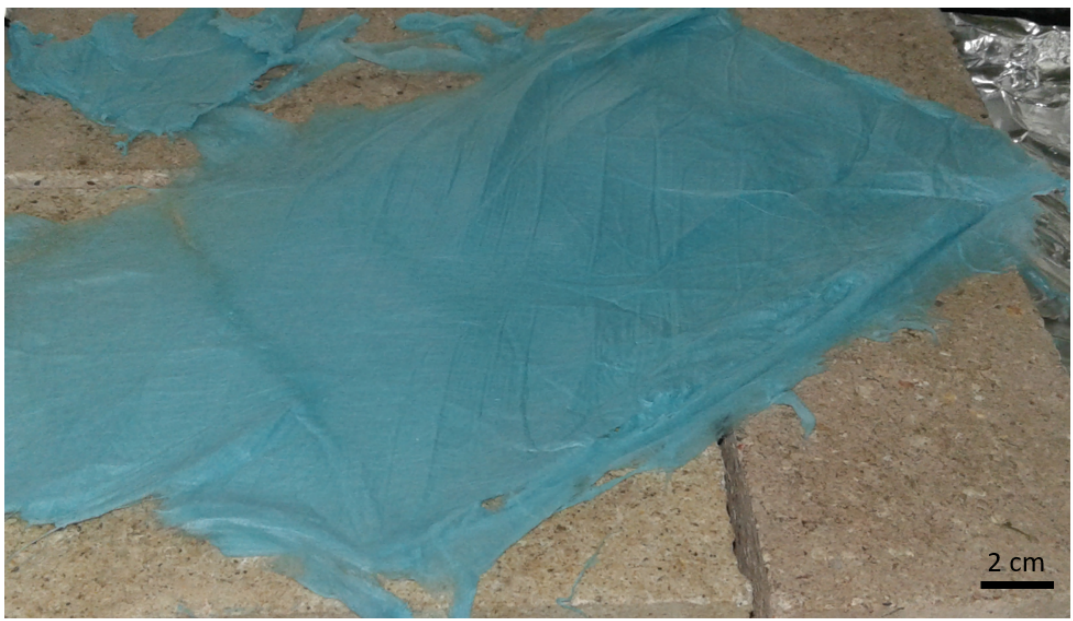

**Figure 3:** Non-woven as-collected polymer sample obtained from the SBS technique.

## 2.5. HEAT TREATMENT

To obtain CuO ceramics, the organics must be eliminated. Thus, the studied samples were heat-treated following three stages. The first stage consisted of maintaining the samples at 200°C for 2h in a conventional electric oven to eliminate possible solvent's excess, avoiding the fusion of the nanowires. The second stage was carried out in a muffle furnace, heating the samples at a rate of 0.5°C/min from room temperature up to 450°C, remaining there for 2h. Such a step results in the organic compounds and the PVP degradation. The third and last stage also used a muffle furnace. The temperature ranged from room temperature to 600°C/2 hours at a rate of 0.5°C/min. After that, non-woven ceramic CuO samples were produced.

## 3. RESULTS AND DISCUSSION

Continuous and homogeneous nanowires are satisfactorily obtained when the polymer's concentration in solution is sufficient to interconnect their long chains with the molecules of the solvent used [23]. This results in a solution with an adequate viscosity (the S51 solution presented a viscosity of 59 cSt, and the S71 was 67.4 cSt) to form fibers using the SBS technique. Therefore, the nanowires' morphology is directly influenced by the concentration of PVP in the precursor solution. However, depending on this concentration, the compressed air's injection rate and pressure need to be adjusted, stabilizing the solution's jet, and forming a continuous non-woven sample without beads [21]. For both S51 and S71, the air pressure was 0.6 bar, and the solution injection rate was 50 µL/min.

After running the appropriate parameters, nanowires were produced, as can be seen in Figure 4. Panel (a) shows S51 where it is possible to verify the desired morphology formation. The fibers are uniform and without beads. By counting the diameters of a hundred fibers, a histogram (shown in Figure 4 (b)) is plotted, and by a gaussian adjustment, an average diameter of (142 ± 50) nm is obtained. Besides that, Figure 4 (c) shows the micrograph of S71. The same uniform nanowire morphology is seen with no beads. The panel (d) histogram shows an average diameter 43% larger than that presented by S51, i.e., (251 ± 60) nm.

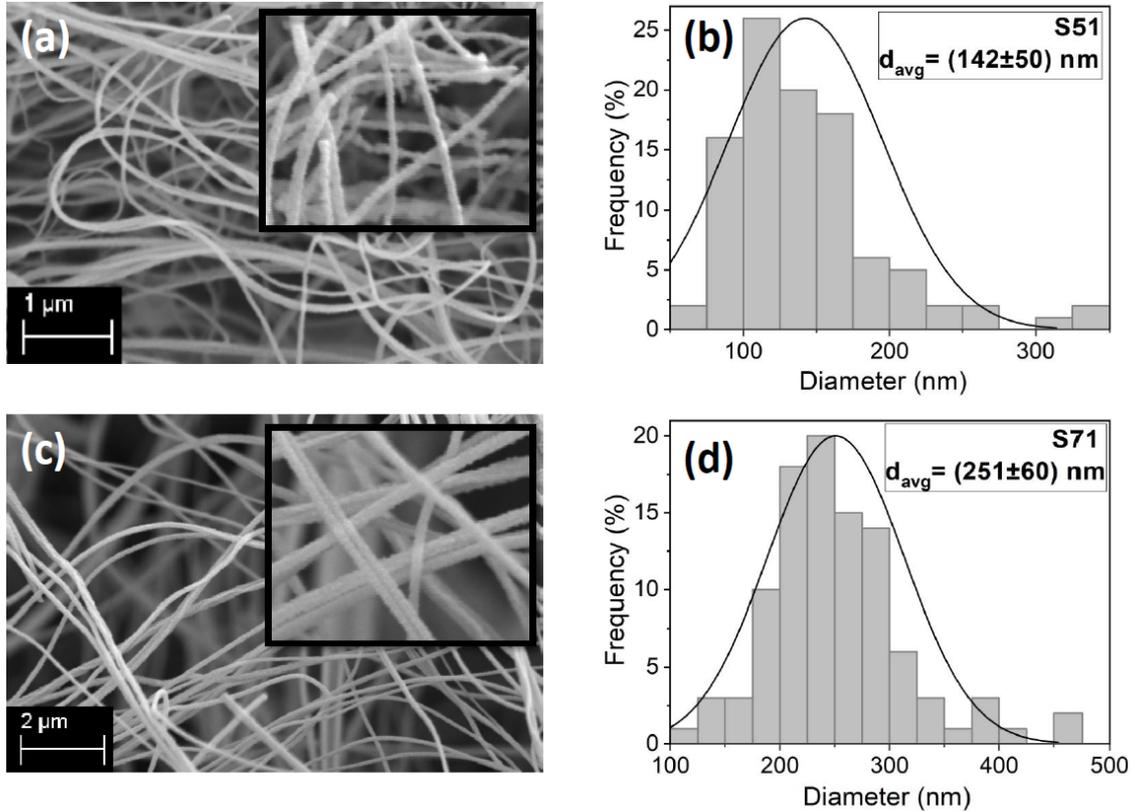

**Figure 4:** Panels (a)/(c) and (b)/(d) show the SEM images and the histograms of S51/S71. Due to the greater precursor solution viscosity, the average diameter of S71 is 43% larger than that presented by S51.

The ceramic nanofibers are not smooth, as can be seen in the insets of Figure 4 (a) and (c). They are formed by small particles giving them a granular morphology. Another important finding is the presence of long wires, however, randomly arranged by the sample.

The XRD diffractogram of the S51 ceramic sample is shown in Figure 5 with the reference data from chart JCPDS 80-1268 for CuO. It can be seen that the desired material is formed without secondary phases, indicating the adequate heat treatment used.

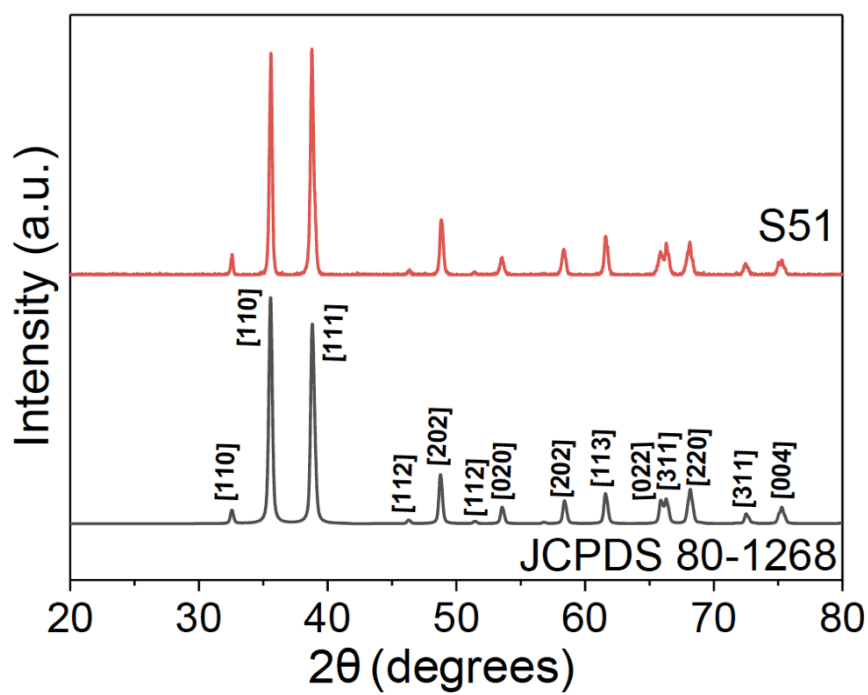

**Figure 5:** S51 XRD diffractogram compared to the reference chart JCPDS 80-1268 after the heat treatment at 600ºC.

Another interesting result is shown in Figure 6. A comparison between the XRD diffractogram of the pure PVP powder, the as-collected sample (before the heat-treatments), and the chart data is shown. The characteristic peaks at 35.5° and 38.8° evidence the formation of CuO without the heat-treatment. The SEM images of the as collected sample are shown in Figure 7. It is impossible to identify possible regions with CuO in the nanowires, which can be due to tiny CuO particles (smaller than the fibers' diameter) spread along the wires.

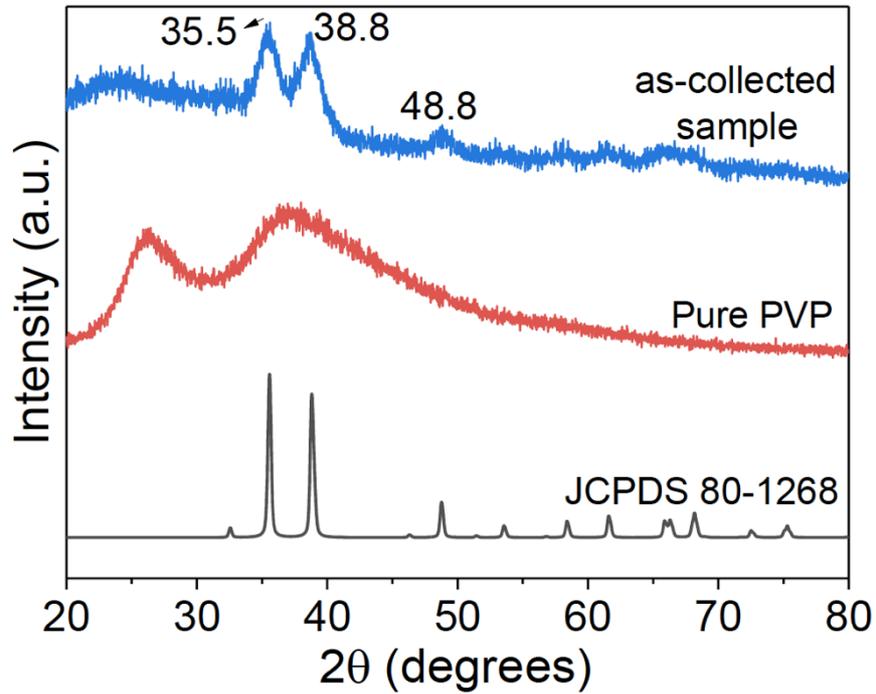

**Figure 6:** XRD diffractograms of the as-collected polymer sample, the pure PVP powder, and the JCPDS 80-1268 chart. The formation of CuO before the heat-treatment processes can be seen.

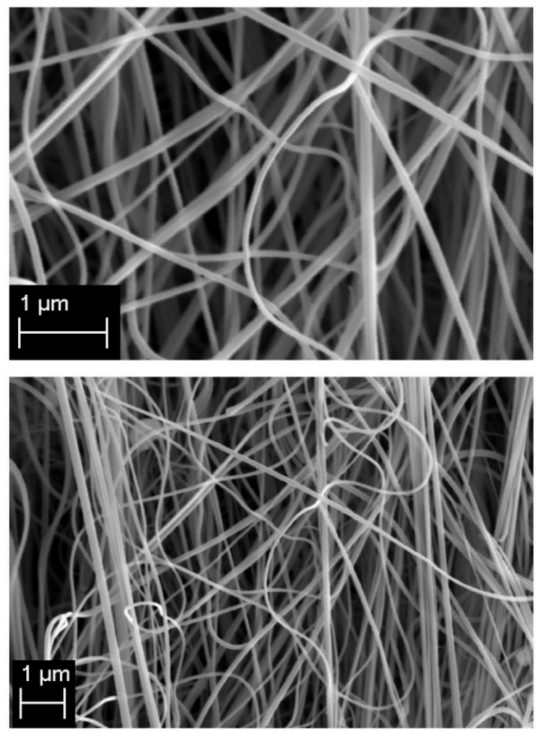

**Figure 7:** SEM image of the as-collected sample. Smooth and randomly distributed wires form the non-woven sample.

To analyze the CuO formation without heat-treatments, a chemical analysis of the precursor solutions, the functional groups of each reagent used, and the interactions between them were investigated. The description of the identified peaks are in Table 2, and the main results are shown in Figure 8. The FTIR results for the seven produced samples are shown in Figure 9.

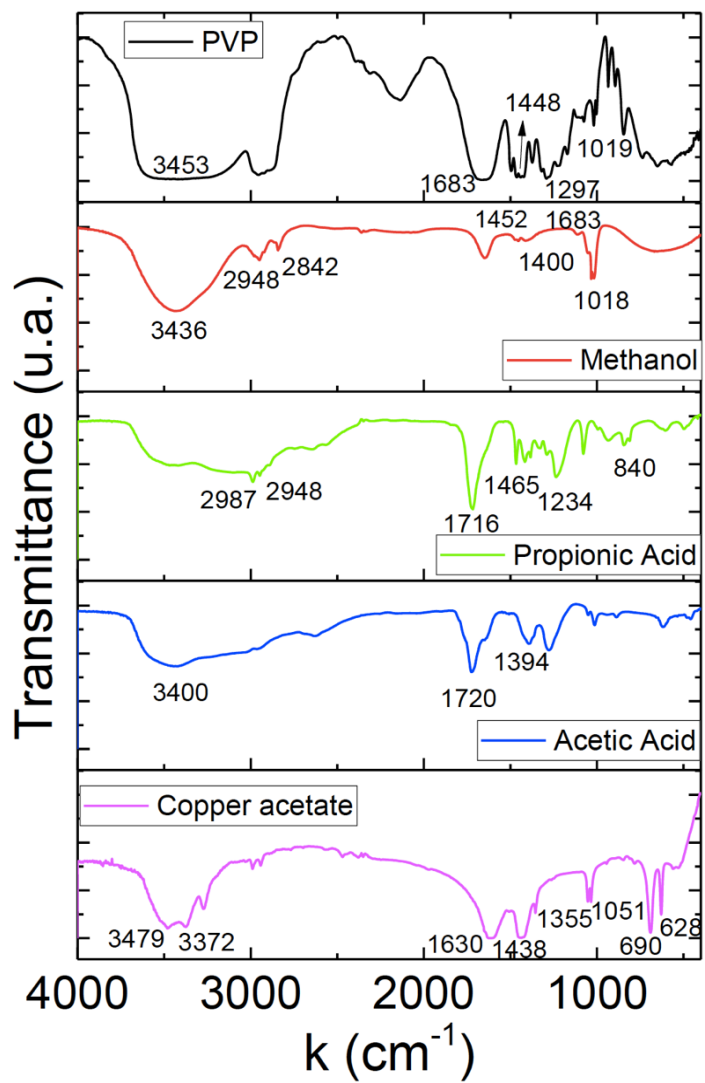

**Figure 8:** FTIR spectrum of each reagent used in the synthesis of the precursor solution.

It is evident from Figure 9 that there are still many peaks related to the solvents, such as in the region of 2983 cm$^{-1}$ peak, referring to propionic acid. For solutions with Ac:PVP ratios from 1:1 to 3:1, such a peak is well evidenced. However, increasing the Ac:PVP ratio in solution, the peak tends to decrease, as can be seen in solutions with ratios from 4:1 to 7:1. Such a relationship is related to the fact that a greater amount of propionic acid is required to release $Cu^{2+}$ ions in the solution while the metallic precursor quantity increases. This fact also means that solutions of lower concentration should not be in a saturated regime. It is still possible to highlight that the peak around 1683 cm$^{-1}$, referring to the PVP amide group, shifted to 1616 cm$^{-1}$ in the produced samples. This displacement occurs when the polymer is complexed to metal ions, thus indicating that in PVP, the coordination of $Cu^{2+}$ cation occurs by forming a bond between the oxygen of the carbonyl group and the copper ions. Thus, it can relate the increase in viscosity with the increase in the ratio Ac:PVP.

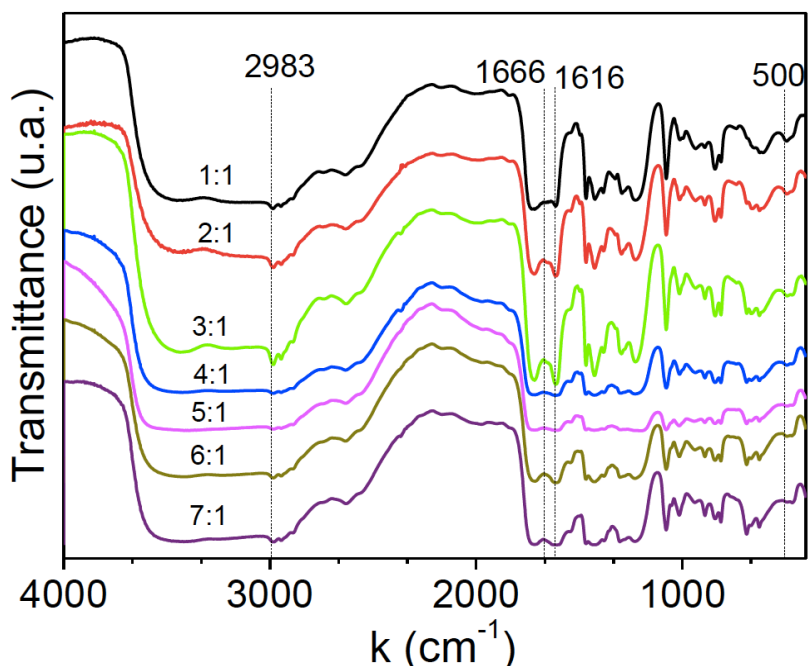

**Figure 9:** FTIR spectra for the seven precursor solutions produced.

The apparent shoulder in the 1666 cm$^{-1}$ region is related to the carbonyl groups of PVP that were not yet complex. The chemical interactions between oxygen and copper (Cu-O) are evidenced by the peak around 500 cm$^{-1}$ [26]. The other peaks not mentioned here refer to the vibrations of other groups, or chemical interactions belonging to the reagents used to prepare the precursor solution.

**Table 2:** FTIR characteristic reagent's peaks used in the precursor solution and their associations.

| Solvent | k (cm$^{-1}$) | Associations |
|---|---|---|
| Copper acetate II | 3479, 3372 | OH unfolding of the water molecule [27] |
| | 1630, 690, 628 | Stretch and angular distortion of the COO$^-$ [28,29] |
| | 1438, 1355, 1051 | Symmetrical and asymmetric stretching and angular distortion of the CH$_3$ [29,30] |
| Acetic acid | 3400 | Stretch of the O-H bond [31] |
| | 1720 | Stretch of C=O [31] |
| | 1394 | Symmetric deformation of CH$_3$ and C-O-H bond [31] |
| Propionic acid | 2987, 840 | Stretch of the O-H bond [32] |
| | 2948 | Stretch of the C-H bond [32] |
| | 1716 | Stretch of C=O [32] |
| | 1465 | Angular deformation C-O-H [32] |
| | 1234 | Stretch of the C-O bond [32] |
| Methanol | 3436 | Stretch of the O-H bond [33,34] |

|     | 2948, 2842 | Stretch of the C-H bond [33,34] |
|     | 1452, 1400 | Asymmetric CH stretch [33,34] |
|     | 1018 | Elongation of the C-O bond [33,34] |
|     | 3453 | N-C bond [35] |
| PVP | 1683 | Amide group (N-C=O) [35] |
|     | 1448, 1297, 1019 | Elongation of C-N and C-N-C bonds [35] |

As a recent methodology applied in this work, a reaction mechanism for synthesizing ceramic oxides using a one-pot-like method is not found in the literature. Therefore, a reaction mechanism for obtaining CuO by the described method is presented here.

Firstly, copper acetate II was used as an ion supplier ($Cu^{+2}$) for the solution. This salt, which is very soluble in the solvents used, has a relatively low commercial cost compared with other copper salts. For the solubilization and dissociation of this salt, propionic acid was used. Glacial acetic acid was added to the reaction system, which in turn had the function of inhibiting the hydrolysis of copper acetate in solution [36]. Methanol entered the process to be the most volatile solvent, which is indispensable when using the SBS technique. The PVP polymer was of fundamental importance for the method since it acted as a stabilizer, and its final OH groups acted as reducers during the CuO/PVP complex synthesis process [20]. This protective effect of PVP occurs through the interaction of the electron pair isolated from the pyrrolidone unit's oxygen and nitrogen atoms. These electrons can be transferred to the metal to form complex ions [37]. The reactions (1), (2), and (3), presented in sequence, refer to the part of obtaining the precursor polymeric solution where, in (1), the interaction of copper ions with the PVP molecules is shown. In (2) is the hydroxylation by the final OH groups of the PVP, forming a complex copper hydroxide with the polymer. Reaction (3) shows the obtaining of the copper oxide complex with PVP – such a mechanism can be confirmed by the XRD data shown in Figure 6, where the formation of CuO in the polymer non-woven sample is evidenced before the heat treatment. Finally, the resulting material at (3) is the same constituent of the as-collected polymer samples. Thus, after the heat-treatments of these samples, the ceramic copper oxide is obtained, as shown in reaction (4).

$$Cu(CH_3COO)_2 \cdot H_2O + PVP \rightarrow Cu(PVP)^{+2} + 2\ CH_3COO^- \quad (1)$$
$$Cu(PVP)^{+2} + 2\ OH^- \rightarrow Cu(OH_2)/PVP \quad (2)$$
$$Cu(OH_2)/PVP \rightarrow CuO/PVP + H_2O \quad (3)$$
$$CuO/PVP \rightarrow CuO \quad (4)$$

## CONCLUSION

CuO nanowires were successfully obtained using the "one-pot" methodology associated with the SBS technique. With the study of the seven produced solutions, it is possible to conclude that the diluted viscosity regime's solutions present difficulties in the blowing process, generating samples with many beads. On the other hand, the concentrated regime solutions are easy to blow, resulting in continuous and homogeneous non-woven samples, as indicated by the SEM micrographs for the samples produced using the Ac:PVP ratios 5:1 and 7:1. Based on FTIR data, the peaks indicate the formation of the CuO/PVP complex from the study of the precursor solutions. Such a fact can also be confirmed when analyzing the XRD of the as-collected polymer samples (before heat treatment). Thus, this work can open new possibilities to use and apply CuO.

**ACKONOWLEDGEMENTS**

We acknowledge the Brazilian agencies São Paulo Research Foundation (FAPESP, grant 2016/12390-6 and grant 2019/06170-1), Coordenação de Aperfeiçoamento de Pessoal de Nível Superior - Brasil (CAPES) - Finance Code 001 and Programa Capes-Print, grant 88887.194785/2018-00, National Council of Scientific and Technological Development (CNPq, grants 302564/2018-7 and 312530/2018-8).

**Author contributions statement**

ANBS: Validation, Investigation, Data Curation, Writing - Original Draft.
MRM: Writing- Reviewing and Editing, Supervision, Validation, Methodology.
RZ: Resources, Writing- Reviewing and Editing, Visualization, Validation, Supervision, Project administration, Funding acquisition, Conceptualization.